\documentclass[12pt,preprint]{aastex}

\newcommand\simlt{\lower.5ex\hbox{$\; \buildrel < \over \sim \;$}}
\newcommand\simgt{\lower.5ex\hbox{$\; \buildrel > \over \sim \;$}}

\begin{document}
\title{Collimation and confinement of magnetic jets by external media}
\author{Amir Levinson$^{1,2}$ and Mitchell C. Begelman$^{3,4}$}
\altaffiltext{1}{School of Physics \& Astronomy, Tel Aviv University, Tel Aviv 69978, Israel; Levinson@wise.tau.ac.il}
\altaffiltext{2}{JILA Visiting Fellow}
\altaffiltext{3}{JILA, University of Colorado and National Institute of Standards and Technology, 440 UCB Boulder, CO 80309-0440, USA; mitch@jila.colorado.edu}
\altaffiltext{4}{Department of Astrophysical \& Planetary Sciences, University of Colorado, 391 UCB, Boulder, CO 80309-0391, USA}

\begin{abstract}
We study the collimation of a highly magnetized jet by a surrounding cocoon that forms as a  result of the interaction of the jet with the external medium.  We show that in regions where the jet is well confined by the cocoon, current-driven instabilities should develop over timescales shorter than the expansion time of the jet's head. We speculate that these instabilities would give rise to complete magnetic field destruction, whereby the jet  undergoes a transition from high-to-low sigma above the collimation zone.   Using this assumption, we construct a self-consistent model for the evolution of the jet-cocoon system in an ambient medium of arbitrary density profile.  We apply the model to jet breakout in long GRBs,  and show that the jet is highly collimated inside the envelope of the progenitor star, and is likely to remain confined well after breakout.  We speculate that this strong confinement may provide a channel for magnetic field conversion in GRB outflows, whereby the hot, low-sigma jet section thereby produced is the source of the photospheric emission observed in many bursts. 
\end{abstract}

\section{Introduction}
The relativistic outflows observed in many compact astrophysical systems are commonly thought to be powered by magnetic extraction of the rotational energy of a neutron star or an accreting black hole.   The energy thereby extracted is transported outward in the form of Poynting flux,  that on large enough scales is converted to kinetic energy flux.    The mechanism by which magnetic energy is converted to kinetic energy has not been identified yet, but it is generally believed to involve gradual acceleration of the flow (e.g., Heyvaerts \& Norman 1989; Chiueh et al. 1991; Bogovalov 1995; Lyubarsky, 2009),   impulsive acceleration (Granot et al. 2011; Lyutikov 2011; Granot 2012), and/or non-ideal MHD effects, specifically  magnetic reconnection (Lyutikov \& Blandford 2003; Giannios \& Spruit 2007; Lyubarsky 2010; McKinney \& Uzdensky 2012).

Astrophysical outflows appear to be highly collimated, and there have been many attempts to explain the observed collimation in different systems.  Understanding the collimation process is important not only from an observational point of view, but also because in ideal MHD flows collimation and acceleration are intimately related (e.g., Begelman \& Li 1994; Vlahakis 2004).   Magnetic fields can cause collimation via magnetic tension.  However, collimation by magnetic tension alone is extremely slow (Eichler 1993; Begelman \& Li 1994; Tomimatsu 1994; Beskin et al. 1998) and cannot account for the inferred collimation scales.  Confinement by the pressure and inertia of an external medium has emerged as a promising alternative (Begelman 1995).    The environments in which astrophysical jets propagate, e.g., accretion disk winds in the case of AGNs, or stellar envelopes in the case of long GRBs, are ideal for this purpose (e.g., Eichler 1983; Levinson \& Eichler 2000; Bromberg \& Levinson 2007, 2009; Kohler et al. 2012).  

The effect of the external medium on the structure of MHD jets has been studied using semi-analytic models (Zakamska et al. 2008; Lyubarsky 2009, 2011; Kohler \& Begelman 2012) and numerical simulations (McKinney \& Blandford 2009; Komissarov et al. 2007; Tchekhovskoy et al. 2010).  However, these studies are restricted to steady state solutions in which the jet boundary is either treated as a rigid wall, or determined by external pressure with a prescribed profile.  In more realistic situations  it is expected that the jet will be  surrounded by a hot cocoon that forms due to side flows of shocked matter from the jet's head, or a nose cone if magnetic pinching is important (Komissarov 1999).  Indeed, numerical simulations (Marti et al. 1997; Aloy et al. 1999; Hughes et al. 2002; Zhang et al. 2003; Morsony et al. 2007; Mizuta \& Aloy 2009; Lazzati et al. 2009) and analytic models (Begelman \& Cioffi 1989; Matzner 2003; Lazzati \& Begelman 2005; Bromberg et al. 2011 (BNPS11)) of purely hydrodynamic jets indicate that under astrophysical conditions anticipated in GRBs, AGNs and microquasars, the surrounding cocoon significantly affects the structure and dynamics of the jet.   Attempts to simulate the propagation of a magnetized jet in an external medium have been limited to two-dimensional Newtonian jets (Clarke et al. 1986; Lind et al. 1989), and relativistic jets with moderate magnetization, $\sigma\simlt 1$, under restricted conditions  (Van Putten 1996; Komissarov 1999). Here $\sigma=B^2/4\pi\rho c^2$, where $B$ and $\rho$ are the proper magnetic field strength and gas density in the jet.  The results of such simulations should be treated with caution, as they cannot account for magnetic field dissipation in the shocked jet and in the nose cone that, as we will argue below, might be important.

In this paper we construct an analytic model for the propagation of a highly magnetized jet in an external medium.  We suggest that a self-consistent treatment of the evolution of the jet-cocoon system may require a proper account of magnetic field dissipation at the jet's head.   Such dissipation is implicitly invoked in our model.  In section 2 we outline the basic model and its key features.  In section 3 we consider the collimation of newly ejected jet material by the surrounding cocoon. In section 4 we compute the evolution of the jet-cocoon system under different conditions.  In section 5 we consider the applications of our model to astrophysical systems.  We conclude in section 6.

\section{\label{model}The basic model}
Consider a magnetized jet propagating along the z-axis in a medium of density $\rho_a(z)$.   We suppose that the jet is injected with a fixed opening angle $\theta_0$, a Lorentz factor $\gamma_j$, and a total power $L_j$.   As the jet pushes its way forward, it accelerates the matter ahead of it and produces  a strong forward shock.  The jet is slowed down behind a reverse shock, to match the velocity of the shocked ambient matter at the contact discontinuity that separates the shocked jet plasma and the shocked ambient gas.   
In the case of a hydrodynamic jet the reverse shock is strong whenever the ambient density is large enough, such that the center of momentum frame moves at a Lorentz factor considerably smaller than $\gamma_j$.   Then, the flow of matter into the jet's head through the forward and reverse shocks is balanced by a sideways flow that feeds a cocoon surrounding the jet.    As long as the cocoon's pressure is sufficiently large, it deflects the streamlines of newly injected fluid and collimates the jet.
As the system evolves, the cocoon expands and its pressure drops, until reaching a level at  which it is too low to confine the jet.   At which point in the course of the evolution this happens depends merely on the density profile of the ambient medium, the injected power $L_j$ and the opening angle $\theta_0$ at the injection point.  
 
 The picture appears to be more involved in the case of  magnetically dominated MHD jets.
In the context of ideal MHD, the reverse shock is expected to be weak in the high sigma limit, in the sense that the compression ratio defined using densities measured in the shock frame is near unity.  This implies that the velocity $u_j=\gamma_j\beta_j$ of the unshocked jet near the head should roughly match the velocity of the head, viz., $u_j\simeq u_h$.   Such a condition requires strong focusing of the section of the jet below the contact discontinuity, that can only be accomplished if the pressure of the confining medium roughly equals the pressure behind the forward shock.  We shall propose in section \ref{sec:mag-cocoon} that this might be accomplished through magnetic stresses, if some fraction of the magnetic field that enters the head through the reverse shock is advected into the cocoon and remains ordered over scales considerably larger than the cross sectional radius of the jet.  An alternative possibility, discussed in section \ref{unmag-cocoon}, is that the 
section of the jet below the contact discontinuity is rendered unstable, thereby giving rise to rapid dissipation of the magnetic field there, and/or entrainment of shocked ambient matter, that can strengthen the reverse shock.  The destruction of the magnetic field below the head may result from  current driven-instabilities (Eichler 1993; Begelman 1998; Mizuno et al. 2009, 2012).  Entrainment of matter may be driven by rapid growth of the Rayleigh-Taylor instability at the contact, which is expected when the head is accelerating (Lyubarsky 2010).   The structure of the cocoon thereby formed consists of an outer part containing the shocked ambient plasma, and an inner part containing the lighter, shocked jet fluid (see figure \ref{fig:schem}). 

As long as the pressure in the cocoon is sufficiently large it will collimate the jet.  The transition from a freely expanding to a confined jet will occur at a radius at which  the 
transverse ram pressure of the conical jet roughly equals the cocoon's pressure.  If the transverse expansion of the unconfined jet is super-magnetosonic, then the streamlines of the jet will be deflected across a  superfast tangential shock.  Otherwise, the collimation will proceed smoothly through the formation of a compression wave that propagates from the jet boundary inwards.  The collimation of the jet by the cocoon is analyzed in section \ref{transition}. The results derived there are used in section \ref{jet-cocoon} to determine the scales over which the cocoon significantly affects the evolution of the jet.

The asymptotic structure of a relativistic, strongly magnetized jet has been analyzed recently by Komissarov et al. (2009) and Lyubarsky (2009, 2011).  Lyubarsky obtained analytic solutions of the transfield equation in the limit $\Omega r \gg 1$ ($r$ is the cylindrical radius), that describe a rigidly rotating, steady jet confined by an external pressure having a power law profile $p_{ext}(z)\propto z^{-\kappa}$.  He examined the behavior of the solutions in the regimes $\kappa<2$ and $\kappa>2$, and showed that 
 when $\kappa<2$  the opening angle of the jet $\theta_j$ decreases continuously such that the jet interior remains in causal contact ($\gamma_j\theta_j\simlt1$) everywhere.  As a consequence,  the jet is accelerated and collimated until it roughly reaches equipartition, where $\sigma\sim1$.  The jet's streamlines have a parabolic shape, $r\propto z^{\kappa/4}$,  with spatial oscillations superimposed on it if the jet is initially out of equilibrium.
 For $\kappa>2$ the jet becomes asymptotically radial, with the final opening angle $\theta_{j\infty}$ depending solely on the pressure profile. In this case $\gamma_{j\infty}\theta_{j\infty}>1$, so that the jet interior is not in causal contact.  The asymptotic Lorentz factor is then practically limited to $\gamma_{j\infty}\simlt \sigma_0^{1/3}\theta_{j\infty}^{-2/3}$, where $\sigma_0\simeq B_0^2/4\pi\rho_0 c^2$ is the value of the magnetization parameter at the injection point.  Since $\gamma_{j\infty}\theta_{j\infty}>1$, the jet remains Poynting dominated in the far zone, viz., $\gamma_{j\infty}<\sigma_0$.  As shown in Lyubarsky (2011), the above results hold not only for a cold jet, but also in the case of a magnetically dominated hot jet. 

The model outlined in the following sections assumes that above the transition zone, at radii where confinement by the cocoon has been communicated to the jet interior, the jet structure can be described by the solution derived in Lyubarsky (2009) for $\kappa<2$. This may be justified in the regime where the propagation of the jet's head is sub- or even mildly relativistic, such that the vertical pressure distribution in the cocoon is sufficiently flat.  As will be shown in section \ref{transition} below, 
the confined jet just upward of the transition region is out of equilibrium, and is expected to oscillate.  
These oscillations may eventually decay to an equilibrium state via production of shocks in the external gas flowing near the jet boundary.   We suppose that near the position of the head the jet is in its equilibrium state, and use the equilibrium solution derived in  Lyubarsky (2009) as a closure condition for the jet-cocoon system analyzed in section \ref{jet-cocoon}.  A schematic illustration of the model described above is presented in figure \ref{fig:schem}.

\section{\label{transition}Transition to the collimation regime}
When the transverse ram pressure of the freely expanding jet roughly equals the cocoon's pressure,  the streamlines of the jet will be deflected until the jet becomes confined by the gas in the cocoon.   As mentioned above, whether the collimation of the jet proceed smoothly or through a formation of a tangential shock, depends on the transverse fast magnetosonic  Mach number of the unconfined jet, $M_{f\perp}=(u_{j}\sin\theta_0)/u_f$, where  $u_f$ is the fast magnetosonic 4-speed, and $\theta_0$ denotes the opening angle of the jet prior to its interaction with the cocoon.  This is confirmed in appendix \ref{sec:appA}, where the jump conditions of an oblique MHD shock are solved.

For a magnetically dominated jet, the  asymptotic Lorentz factor is limited to $\gamma_j\simlt(\sigma_0/\theta_0^2)^{1/3}$ (see discussion in the preceding section).  If the jet is injected cold, then  $\gamma_j\sigma_j=\sigma_0$ to $O(\sigma_j^{-1})$ (see Equation (\ref{enthalpy-cons})), yielding $\gamma_j\beta_j\theta_0/\sqrt{\sigma_j}\simlt1$ in the asymptotic limit.  Consequently, for  a cold jet the transverse expansion is marginally sub-magnetosonic and we expect smooth collimation.  The confinement of the jet in this case is communicated to the inner regions by a fast magnetosonic wave that propagates from the jet boundary inwards (Kohler \& Begelman 2012). If the jet is injected hot then it may become superfast when interacting with the cocoon, and a tangential collimation shock will form.   At the tangential contact discontinuity the pressure of the shocked jet layer must be equal to the cocoon's pressure $p_c$. 
Then, assuming that the transverse momentum flux is roughly uniform across the shocked layer we obtain, using Equation (\ref{app-eq-momen}),
\begin{equation}
p_c=(h_j+\sigma_j)\rho_jc^2\gamma_j^2\beta_j^2\cos^2\psi_j
\end{equation}
for a superfast flow, viz., $M_{f\perp}\gg 1$.   Here $\psi_j$ is the angle between the fluid velocity and the shock normal (see appendix \ref{sec:appA}).

Adopting $p_c=p_0(z/R_L)^{-\kappa}$ for the external pressure, $R_L$ being the radius of the light cylinder, and recalling that $L_j=(h_j+\sigma_j)\rho_j\gamma_j^2\beta_jc^3\pi \tan^2\theta_0 z^2$, yields
\begin{equation}
\cos\psi_j=A\tan\theta_0(z/R_L)^{1-\kappa/2},\label{cos-psi-1}
\end{equation}
with $A=(\pi c p_0R_L^2/L_j)^{1/2}\simeq(4\pi p_0/B_0^2)^{1/2}$, where $B_0$ is the characteristic magnetic field  at the light cylinder.  We see that this condition is essentially the same as in the hydrodynamic case (Bromberg \& Levinson 2009; Van Putten \& Levinson 2012), as naively expected.   In the small angle approximation $\cos\psi_j=r_s/z-dr_s/dz$, where $r_s(z)$ is the cylindrical radius of the shock front.   Substituting the latter expansion  into Equation(\ref{cos-psi-1}) we obtain 
\begin{equation}
\frac{dr_s}{dz}=\frac{r_s}{z}-A\tan\theta_0(z/R_L)^{1-\kappa/2}.
\label{ODE-evl}
\end{equation}
Solving Equation (\ref{ODE-evl}), subject to $r_s(z=R_L)=R_L\tan\theta_0$, gives
\begin{equation}
r_s(z)=z\tan\theta_0-\frac{2A}{2-\kappa}z\tan\theta_0[(z/R_L)^{1-\kappa/2}-1].
\label{shc-evl}
\end{equation}
 The point $z^\star$ at which the shock reaches the axis is determined from the condition $r_s(z=z^\star)=0$:
\begin{equation}
z^\star=R_L\left(\frac{2-\kappa}{2A}+1\right)^{1/(1-\kappa/2)}.
\label{zstar}
\end{equation}
The latter result is used in section \ref{unmag-cocoon} to determine the scales over which collimation of the jet by the cocoon occurs (Equation (\ref{z_c})).

If the motion of the head is slow enough, the cocoon is anticipated to be roughly isobaric. Then $\kappa=0$, and if $A \ll 1$ we have $z^\star/R_L\simeq (B_0^2/4\pi p_0)^{1/2}$.   The radius of the jet at this distance is $R_j^\star=R_L\tan\theta_0 (B_0^2/4\pi p_0)^{1/2}$.   The radius of an equilibrium cylinder is $R_e=R_L(B_0^2/2\pi p_0)^{1/4}$ (Lyubarski 2009).    Our analysis assumes that $\gamma_j\sin\theta_0>\sqrt{\sigma_j/h_j}>1$.  Since $\gamma_j\simlt R_j/R_L$, it implies that $\sin\theta_0>\sqrt{\sigma_j/h_j}(R^\star_j/R_L)^{-1}$.  At small angles $\sin\theta_0\simeq\tan\theta_0=\sqrt{2}R_j^\star R_L/R_e^2$, and the above results yield $R^\star_j\simgt (\sigma_j/2h_j)^{1/4}R_e > R_e$.  If $\tan\theta_0<\sqrt{\sigma_j/h_j} (R_e/R_L)^{-1}$ it means that the transverse expansion of the jet is sub-magnetosonic, so that collimation proceeds smoothly. 

Because $R_j>R_e$ at $z^\star$, we expect spatial oscillations of the jet to ensue above the transition region.  To illustrate this, consider the propagation of a jet in a confining medium having a uniform pressure, $p_c=p_0=$ const.   Suppose that at $z=z_0$ the radius of the jet satisfies $R_j=R_0$ and $dR_j/dz=0$. The solution of the transfield equation in this case reduces to (Lyubarsky 2009)
\begin{equation}
R_j(z)=R_0[\cos^2\{\sqrt{3/2}A(z-z_0)\}+(R_e/R_0)^4\sin^2\{\sqrt{3/2} A(z-z_0)\}]^{1/2},
\end{equation}
with $A$ and $R_e$ as defined above.  For $R_0=R_e$ the latter equation yields $R_j(z)=R_e=$ const. For any other values the jet oscillates. Such oscillations are present  for any pressure profile with $\kappa\le2$, when the initial jet radius deviates from the equilibrium value.  These spatial oscillations may eventually decay and the jet radius will approach $R_e$.

%%%%%%%%%
\section{\label{jet-cocoon}Evolution of the jet-cocoon system}
The temporal evolution of the jet-cocoon system is determined by the density profile of the ambient medium and the parameters of the injected jet, assumed to be given.
As explained in section \ref{model} above, matching of the jet and head Lorentz factors can be accomplished through magnetic pinching or, alternatively, non-ideal MHD effects that allow the formation of  a strong reverse shock below the contact discontinuity.   The evolution of the system in the former case is explored in section \ref{sec:mag-cocoon}, and in the latter case in section \ref{unmag-cocoon}.

The energy momentum tensor of the unshocked jet can be expressed as
\begin{equation}
T_j^{\mu\nu}=(w_j+b_j^2)u_j^\mu u_j^\nu+(p_j+b_j^2/2)g^{\mu\nu}-b_j^\mu b_j^\nu\label{Tmunu_j}
\end{equation}
where $p_j$ and $w_j$ are the pressure and specific enthalpy, respectively, $u_j^\mu$ is the 4-velocity, and $\sqrt{4\pi}b_{\mu}=F^\star_{\mu\nu}u^\nu$ is the  magnetic field vector,  $F^\star$ being the dual electromagnetic tensor.  
 Since well above the light cylinder the azimuthal magnetic field of the unshocked jet satisfies $B_\phi\simeq r\Omega B_p \gg B_p$, where $r$ is the cylindrical radius of the magnetic surface $\Psi(r,z)$ and $\Omega(\Psi)$ its angular velocity, we can neglect the poloidal field $B_p$.  Then $b_j^\mu=(0,0,b_j,0)$, where $\sqrt{4\pi}b_j=B_{j\phi}/\gamma_j$ is the proper magnetic field.    The jet power is obtained upon integration of the energy flux from the jet axis to its boundary, at $r=R_j$:
 \begin{equation}
 L_j=\int_0^{R_j}T_j^{0z} 2\pi rdr=\int_0^{R_j}(w_j+b_j^2)\gamma_j^2c\beta_j2 \pi rdr.
 \end{equation}
We suppose that the structure of the confined jet well above the transition region can be approximated by the equilibrium solution obtained in Lyubarsky (2009) in the limit of rigid rotation,  $\Omega=$ const.  Then, $L_j\simeq (B_0^2/4\pi)c\pi R_L^2$, where $B_0= \Psi_0/R_L^2$ is the characteristic magnetic field at the light cylinder, and $2\pi\Psi_0\equiv 2\pi \Psi(R_j,z)$ is the total magnetic flux subtended by the jet.
In terms of the cocoon's pressure at the jet boundary, $p_c(z)$, the cross-sectional radius of the  confined jet is given by (Lyubarsky 2009)
\begin{equation}
R_j/R_L\simeq \left(\frac{B_0^2}{2\pi p_c}\right)^{1/4}\simeq\left(\frac{2L_j}{\pi cR_L^2p_c}\right)^{1/4} ,\label{Rj}
\end{equation}
and the Lorentz factor by $\gamma_j\simeq R_j/R_L$.
 
Let $\beta_h$ denotes the velocity of the contact discontinuity (the head) and $\gamma_h$ the corresponding Lorentz factor.  Momentum balance at the forward and reverse shocks gives 
\begin{equation}
\int_0^{R_j}[(w_j+b_j^2)\gamma_j^2\gamma_h^2(\beta_j-\beta_h)^2+p_j+b_j^2/2]2\pi rdr=\int_0^{R_f}[w_a\gamma^2_h\beta_h^2+p_a]2\pi rdr,\label{momentum-balance}
\end{equation}
where $w_a$ and $p_a$ are the enthalpy and pressure of the ambient medium, respectively, and $R_f$ is the cross sectional radius of the forward shock.  For simplicity, we shall consider a cold medium, $p_a=0$, and assume that the effective cross sectional radius of the forward shock roughly equals that of the jet.  Then, Equation (\ref{momentum-balance}) yields:
$l\gamma_h^2\beta_j(1-\beta_h/\beta_j)^2+\tilde{p}=\gamma_h^2\beta_h^2$, in terms of the dimensionless parameters,
\begin{eqnarray}
 l=L_j/(\pi R_j^2\rho_ac^3),\label{tilde(L)}\\
 \tilde{p}=(p_j+b_j^2/2)/\rho_ac^2.  \label{tilde(p)}
 \end{eqnarray}
Here, $l$ represents the ratio between the total energy density of the jet, $L_j/c\pi R_j^2$,  and the rest-mass energy density of the surrounding matter, as measured in the rest frame of the ambient medium.  Likewise, $\tilde{p}$ is the ratio between the total jet pressure and the rest-mass energy density of the ambient matter.  The solution for the head velocity reads:
\begin{equation}
\beta_h=\beta_j\frac{l-[l^2-(l+\tilde{p}/\beta_j^{2})(l-\tilde{p}-1)]^{1/2}}{l-\tilde{p}-1}.\label{beta_h-general}
\end{equation}
Note that for a highly magnetized jet, $b_j^2 \gg w_j$, one has, to a good approximation, $\tilde{p}=l/(2\gamma_j^2)$.
When the reverse shock is strong $l \gg \tilde{p}$ and the latter solution simplifies to 
\begin{equation}
\beta_h=\frac{\beta_j}{1+l^{-1/2}},\label{beta_h-L}
\end{equation}
 as derived in BNPS11.  When the reverse shock is weak or absent, specifically when $\gamma_j^2\gamma_h^2(\beta_j-\beta_h)^2<1$, Equation (\ref{beta_h-general}) reduces to
 \begin{equation}
\beta_h=\sqrt{\frac{\tilde{p}}{1+\tilde{p}}}.
\end{equation}

In order to proceed, we need to specify the conditions in the inner cocoon.  In what follows, we consider two different scenarios. 
%%%%%%%%%
\subsection{\label{sec:mag-cocoon}Magnetized cocoon}
In this section we consider the possibility that the jet is pinched by magnetic stresses in the inner cocoon, assuming that a fraction $\xi_B$ of the 
toroidal magnetic field (more precisely, the fraction of magnetic flux) that enters the head through the reverse shock is advected into the inner cocoon.  
Suppose that the rate of separation of the forward and reverse shocks is much slower than the velocity of the head; that is, $\beta_h\simeq\beta_r$.  Then, the rate at which magnetic flux is added to the head, as measured in the star frame, is  roughly 
\begin{equation}
\frac{d\Phi_{Bj}}{dt}\simeq(\beta_j-\beta_h)\int_{0}^{R_j}B_{j\phi}(r)dr=B_{j\phi}(R_j)R_j(\beta_j-\beta_h)/2, 
\end{equation}
assuming a uniform current density inside the jet.  Here, $B_{j\phi}(R_j)$ denotes the value of $B_{j\phi}$ at the jet boundary $r=R_j$.  For illustration, we assume that the return current is flowing in a thin sheet  at some radius $R_c>R_j$.  Then,  the magnetic field in the inner cocoon scales as $B_c\propto r^{-1}$ with cylindrical radius $r$.   The rate at which magnetic flux is added to the cocoon is approximately $d\Phi_{Bc}/dt\simeq\beta_h\int_{R_j}^{R_c}B_c dr\simeq\beta_h R_jB_{c0}\ln(R_c/R_j)$, where $B_{c0}$ is the value of $B_c$ at the jet boundary, $r=R_j$.  Flux conservation, viz., $d\Phi_{Bc}/dt=\xi_Bd\Phi_{Bj}/dt$,  implies 
\begin{equation}
B_{c0}=(\xi_B/2)B_{j\phi}(R_j)(\beta_j/\beta_h-1)[\ln(R_c/R_j)]^{-1}.\label{B_c0}
\end{equation}
In order not to crush the jet, the cocoon's pressure at the jet boundary, $p_{c0}\simeq B_{c0}^2/8\pi$, should not exceed the jet pressure, $p_j\simeq[B_{j\phi}(R_j)]^2/(8\pi \gamma_j^2)$.  This yields the condition
\begin{equation}
\ln(R_c/R_j)>(\xi_B/2)\gamma_j(\beta_j/\beta_h-1).\label{Rc/Rj}
\end{equation}
Since $\ln(R_c/R_j)=$ a few, Equation (\ref{Rc/Rj}) implies $\xi_B(\beta_j-\beta_h) \ll 1$ in the relativistic regime $\gamma_j \gg 1$.   Unless the fraction $\xi_B$ is very small, the latter condition means that $|\beta_j-\beta_h| \ll 1$.\footnote{For a uniform current distribution in the cocoon with $B_c(R_c)=0$ we obtain $d\Phi_{Bc}/dt\simeq\beta_h R_jB_{c0}[\ln(R_c/R_j)/(1-R_j^2/R_c^2)-1/2]$, suggesting that the latter condition is quite robust.}

To illustrate some key features of jet focusing by the inner cocoon, we compute the evolution of the system by invoking the extreme condition $\gamma_h=\gamma_j$.   This, of course, is a gross approximation, as some velocity difference is required in order that magnetic flux will be advected into the inner cocoon, as indicated by Equation (\ref{B_c0}). Nonetheless, for $\beta_j/\beta_h-1\simlt (\gamma_j\gamma_h)^{-1}$ it may represent a reasonable approximation of a more realistic situation.  Furthermore, we assume that the inner jet can be described by the equilibrium solution derived in Lyubarsky (2009), so that $\gamma_j\simeq R_j/R_L$.  With $\beta_j=\beta_h$, Equation (\ref{momentum-balance}) yields $\gamma_h^2=\tilde{p}\simeq l/2\gamma_j^2$, and since $\gamma_h=\gamma_j$ we have, using Equation (\ref{tilde(L)}),
\begin{equation}
\gamma_h=R_j/R_L=\left(\frac{L_j}{2\pi R_L^2 \rho_0c^3}\right)^{1/6}\tilde{z_h}^{\alpha/6}\label{gamma_h-mag}
\end{equation}
for a density profile $\rho_a=\rho_0\tilde{z_h}^{-\alpha}$, where $\tilde{z}_h=z_h/R_L$.   Comparing the latter with the unmagnetized case discussed in the next section (Eq. (\ref{gamma_h-unmag}) with $\eta\simeq1$), it is seen that the head Lorentz factor is larger by a factor $1.5\tilde{z}_h^{1/9}$ when magnetic pinching is effective.  We remind the reader that the above results assume that the radius of the forward shock roughly equals the jet radius  at the head.  If magnetic pinching is effective mainly at a nose surrounding the head it would mean that the jet is strongly focused only in the vicinity of the head. Then the assumption that the radius of the forward shock matches that of the head is questionable.  

Let us proceed by assuming that the outer cocoon is roughly isobaric, has a cylindrical geometry, and contains radiation-dominated gas, as described in detail in \S \ref{unmag-cocoon}. Then, its pressure can be approximated  by Equation (\ref{pc-1}) below.  Using  Equations (\ref{gamma_h-mag}) and (\ref{pc-1})  with $\beta_c^2=p_c/\rho_ac^2$ and  $\beta_h^{-1}-1\simeq 1/2\gamma_h^2$, we arrive at
\begin{equation}
p_c=(\eta/3)^{1/2}\rho_ac^2\gamma_h^2\tilde{t}^{-1}\propto\tilde{z_h}^{-(3+2\alpha)/3}.\label{pc-mag}
\end{equation}
Under the assumption  $B_c\propto r^{-1}$, the pressure exerted on the jet by the inner cocoon is related to the thermal pressure $p_c$ of the outer 
cocoon, at $r=R_c$, through 
\begin{equation}
p_{c0}=p_c(R_c/R_j)^2.
\end{equation}
Now, for the equilibrium solution adopted here the jet radius $R_j$ is given by Equation (\ref{Rj}) with $p_c$ replaced by $p_{c0}$, specifically $R_j/R_L=(2L_j/\pi c R_L^2p_{c0})^{1/4}=(2L_j/\pi c R_L^2p_{c})^{1/4}(R_c/R_j)^{1/2}$.  Combining the latter relation with Eqs. (\ref{gamma_h-mag}) and (\ref{pc-mag}) gives the radius of the current sheet in the inner cocoon:
\begin{equation}
R_c/R_j=2(3/\eta)^{1/4}\tilde{t}^{1/2}.
\end{equation}

The above derivation implicitly assumes that in the vicinity of the head the relativistic jet is focused by magnetic pinching to a level at which the jet Lorentz factor can be continuously matched to the head.   One might naively conclude that if the ambient medium is sufficiently dense, then  
ultimately a bubble of submagnetosonic, high Poynting flux material will fill the expanding cocoon, quenching the inner jet to a narrow cylinder of radius  $R_j\sim R_L$ that propagates outwards at a subrelativistic speed, $\gamma_j\simeq R_j/R_L\simeq 1$.  
In reality, such a structure is expected to be extremely unstable.  What seems likely to happen is that current-driven instabilities will destroy the magnetic field in the inner jet before it is even deposited in the cocoon.   Since the comoving growth time of the instability roughly equals the Alfv\'en crossing time of the jet, $t^\prime\sim R_j/v_A$ (Begelman 1998), the length scale over which the instability develops is $\lambda_{CD}\sim \gamma_jct^\prime\sim \gamma_j R_j$. If the jet is well collimated by the cocoon, then this scale is generally smaller than 
the distance $z_h$ of the jet's head from the origin.   It is, therefore, conceivable that the magnetization of the jet in some region between the head and the collimation zone is strongly reduced by the instability.  We examine the consequences of such a process next.

%%%%%%
\subsection{\label{unmag-cocoon}Unmagnetized cocoon}
If the magnetic field dissipates before reaching the cocoon, as argued above, then magnetic stresses in the inner cocoon can be ignored.   As shown below, in the absence of magnetic pinching the jet cannot be sufficiently focused, and its Lorentz factor $\gamma_j$  may be much larger than the Lorentz factor $\gamma_h$ of the head, as in the pure hydrodynamic case.   This means that the jet must decelerate across a strong reverse shock.
 We envision that the shock is formed in a low-sigma section of the jet, following the destruction of the incident magnetic field by the instabilities described above.  

Under the conditions envisaged the pressure in the cocoon is anticipated to be radiation-dominated, thus we adopt an adiabatic index of $4/3$.   Following BNPS11, we assume that the cocoon is roughly isobaric, and approximate its geometry as a cylinder of height $z_h=c\int\beta_hdt$ and cylindrical radius $r_c=c\int\beta_cdt$, where $\beta_c=(p_c/\rho_ac^2)^{1/2}$ is the lateral expansion velocity of the outer cocoon.  The former assumption, that the energy distribution in the cocoon is approximately uniform, introduces a considerable simplification and may be justified  when the motion of head is sufficiently slow. With  the above approximations the cocoon's pressure is given by  $p_c=E_c/3V_c$, where $E_c=\eta L_j\int(1-\beta_h)dt$ is the total energy deposited in the cocoon 
and $V_c=\pi r_c^2z_h$ its volume.  
The parameter $\eta$ represents the fraction of the energy that enters the cocoon, as explained in BNPS11; at sufficiently low Lorentz factors of the head, $\gamma_h<2z_h/R_j$, for which it is in causal contact across its transverse direction, $\eta=1$.   Otherwise $\eta=2z_h/(\gamma_hR_j)$. 
Taking for simplicity $r_c=c\int\beta_cdt\simeq \beta_cc t$, and likewise $z_h=c\beta_ht$, we obtain
\begin{equation}
p_c=\frac{\eta L_j}{3\pi cR_L^2}\frac{(\beta_h^{-1}-1)}{\beta_c^2\tilde{t}^{2}},\label{pc-1}
\end{equation}
in terms of the fiducial time $\tilde{t}=ct/R_L$. This expression is accurate up to an order unity factor that  depends on the density profile of the ambient medium (see BNPS11).   In the regime where the reverse shock is strong,  $\beta_h$ can be approximated by Eq. (\ref{beta_h-L}). 
Substituting $\beta_c$ and $\beta_h$ into the last equation and solving for $p_c$ yields 
\begin{equation}
p_c=\left(\frac{\eta L_j\rho_ac}{3\pi R_L^2}\right)^{1/2}l^{-1/4}\tilde{t}^{-1}.\label{pc-2}
\end{equation}

Next, we suppose that near the head the cross-sectional radius of the jet can be approximated by Equation (\ref{Rj})
and the Lorentz factor by $\gamma_j=R_j/R_L$.   %We envision that the matching of the high sigma jet at the head involves the destruction of magnetic field by the instabilities described above.
Solving Equations (\ref{Rj}), (\ref{tilde(L)}) and (\ref{pc-2}) one finds
\begin{eqnarray}
l=0.26\eta^{2/9}\left(\frac{L_j}{R_L^2\rho_ac^3}\right)^{2/3} \tilde{t}^{-4/9}=l_0\tilde{z}_h^{2\alpha/3}\tilde{t}^{-4/9},\label{l-2}
%R_j/R_L=(12/\eta)^{1/6}l^{1/4}\tilde{t}^{1/3}=1.5\eta^{-1/6}l_0^{1/4}\tilde{z}^{\alpha/6}\tilde{t}^{2/9}
\end{eqnarray}
where we adopt an ambient density profile of the form $\rho_a(z)=\rho_0\tilde{z}^{-\alpha}$, with $\tilde{z}_h=z_h/R_L$, and define
\begin{equation}
l_0=0.26\eta^{2/9}\left(\frac{ L_j}{R_L^2\rho_0c^3}\right)^{2/3} .\label{l_0}
\end{equation}
The radius and Lorentz factor of the jet are found from (\ref{Rj}),  (\ref{pc-2}) and (\ref{l-2}):
\begin{equation}
\gamma_j=R_j/R_L=(12/\eta)^{1/6}l^{1/4}\tilde{t}^{1/3}=1.5\eta^{-1/6}l_0^{1/4}\tilde{z}_h^{\alpha/6}\tilde{t}^{2/9}.\label{Rj-2}
\end{equation}
Result (\ref{Rj-2}) holds at times at which $\gamma_j<\gamma_{max}$. 

The position of the head at time $\tilde{t}$ is given by  $\tilde{z}_h=\beta_h \tilde{t}$. From Eq. (\ref{beta_h-L}) it is readily seen that if $l \ll 1$ then $\tilde{z}_h\sim l^{1/2}\tilde{t}$, and if $l \gg 1$, $\tilde{z}_h\sim \tilde{t}$.  In the former case we have
\begin{equation}
l=l_0^{9/7} \tilde{z}_h^{(6\alpha-4)/7},\label{L-scale1}
\end{equation}
and in the latter case ($l \gg 1$)
\begin{equation}
l=l_0 \tilde{z}_h^{(6\alpha-4)/9}.\label{L-scale2}
\end{equation}
Consequently, for $\alpha>2/3$, $l$ increases with $z_h$ and the head accelerates.   This is different than the hydrodynamic case where this happens at $\alpha>2$ (c.f., BNPS11).   
Note that when the head's motion becomes relativistic the Lorentz factor of the head is given by
\begin{equation}
\gamma_h\simeq l^{1/4}/\sqrt{2}=0.68\eta^{1/18}\left(\frac{L_j}{2\pi R_L^2 \rho_0c^3}\right)^{1/6} \tilde{z}_h^{(3\alpha-2)/18},\label{gamma_h-unmag}
\end{equation}
where Eq. [\ref{beta_h-L}] has been employed.
The pressure inside the cocoon satisfies $p_c\propto z_h^{-s}$ with $s=(2\alpha+8)/7$ when $l \ll 1$ and $s=(6\alpha+8)/9$ when $l \gg 1$.  
From Eq. (\ref{Rj}) and the relation $\gamma_j=R_j/R_L$ we have $\gamma_j\propto \tilde{z}^{s/4}=\tilde{z}^{(3\alpha+4)/18}$ for $l \gg 1$.  The scaling of $\gamma_h$ and $\gamma_j$ with $z_h$ confirms that as long as the jet is confined by the cocoon it always accelerates faster than the head.
The jet is considered collimated as long as $z^\star<z_h$, where $z^\star$ is given by Eq. (\ref{zstar}) with $\kappa=0$ for an isobaric cocoon.   The jet will be unconfined only when  $A \ll 1$,  for which $\tilde{z}^{\star}\simeq (L_j/\pi cR_L^2p_c)^{1/2}\propto \tilde{z}_h^{s/2}$.  When $s<2$ ($\alpha<5/3$) the head advances faster than $z^\star$, and the jet becomes confined at distances $z>z_c$, where 
\begin{equation}
\tilde{z}_c=(1.6\eta^{-2/9}l_0^{1/2})^{9/(5-3\alpha)}.\label{z_c}
\end{equation} 
When $s>2$ ($\alpha>5/3$), the jet is confined at $z<z_c$, and becomes unconfined at $z>z_c$. 

The model presented above implicitly assumes that current-driven instabilities lead to magnetic field dissipation above the transition zone, at $z>z^\star$.  This assumption is justified provided the instability growth time is shorter than the dynamical time for jet fluid to reach the head.  As argued at the end of section \ref{sec:mag-cocoon}, the instability growth length is  $\lambda_{CD}\sim \gamma_j R_j\beta_{A}^{-1}\simeq R_j^2/(\beta_AR_L)$, where $\beta_A$ is the Alfv\'en speed in units of c, for which  Equation (\ref{Rj}) yields $\lambda_{CD}/R_L\simeq (2L_j/\pi cR_L^2p_c)^{1/2}\beta_A^{-1}=\sqrt{2}\beta_A^{-1}\tilde{z}^\star$.  Now, as long as the jet is confined by the cocoon $z^\star<z_h$, hence the magnetic field in the jet has sufficient time to dissipate before it reaches the head provided $\beta_A>z^\star/z_h$.  In particular, in the regime where the head is sub-relativistic we find $\lambda_{CD}/z_h\sim\beta_A^{-1}\tilde{t}^{-1/3}\sim\beta_A^{-1}l_0^{3/14}\tilde{z}_h^{(\alpha-3)/7}$, and in the regime where the head is transrelativistic  $\lambda_{CD}/z_h\sim\beta_A^{-1}l_0^{1/2}\tilde{z}_h^{(3\alpha-5)/9}$.

It is worth noting that in a relativistically hot, pure hydrodynamic flow the cross-sectional radius and Lorentz factor scale as $\gamma_j\propto R_j\propto p_j^{-1/4}$ with pressure $p_j$. By comparing with Equation (\ref{Rj}) it is seen that conversion of magnetic energy to kinetic energy in the confinement region does not change the scaling of the outflow parameters, so that the use of Equation (\ref{Rj}) is justified even if the jet becomes kinetic dominated in the vicinity of the head.  
 
%%%%%%%%%%%%%%
\section{Applications}
%\subsection{GRBs}
We examine first the application of the above results to GRBs, assuming an unmagnetized cocoon.  
In the context of the collapsar scenario for long GRBs, the jet propagates inside the envelope of a massive star before breaking out to produce the observed signal.  For illustration, we invoke a WR star of mass $M\sim10 M_{\sun}$ and radius $R_\star\sim R_\sun$.  The density profile in the stellar envelope may be expressed as $\rho_a(z)=\bar{\rho}(z/R_\star)^{-\alpha}$, with $\alpha<3$, where the average density is roughly $\bar{\rho}\simeq 5(3-\alpha)(M/10M_\sun)(R_\star/R_\sun)^{-3}$ g cm$^{-3}$.  From Equation (\ref{l_0}) we obtain
\begin{equation}
l_0=2\times10^3 \eta^{2/9}\left(\frac{R_L}{R_\star}\right)^{2\alpha/3}\left(\frac{L_{j52}}{R_{L7}^2}\right)^{2/3}\left[\frac{(3-\alpha)M}{10\ M_\sun}\right]^{-2/3}\left(\frac{R_\star}{R_\sun}\right)^2
\end{equation}
where  $R_L=10^7R_{L7}$ cm.  For $\alpha=2$ and the above choice of $M$, $R_\star$ and $R_L$ we have, using (\ref{L-scale1}) with $\eta=1$,
\begin{equation}
l\simeq5\times10^{-3}\left(\frac{L_{j52}}{R_{L7}^2}\right)^{6/7}\tilde{z}_h^{8/7}\simeq123
\left(\frac{L_{j52}}{R_{L7}^2}\right)^{6/7}\left(\frac{z_h}{R_\star}\right)^{8/7}
\end{equation}
in the sub-relativistic regime, and from (\ref{L-scale2})
\begin{equation}
l\simeq1.5\times10^{-2}\eta^{2/9}\left(\frac{L_{j52}}{R_{L7}^2}\right)^{2/3}\tilde{z}_h^{8/9}\simeq
40\eta^{2/9}\left(\frac{L_{j52}}{R_{L7}^2}\right)^{2/3}\left(\frac{z_h}{R_\star}\right)^{8/9}
\end{equation}
in the relativistic regime ($l \gg 1$).  The Lorentz factor of the head can be expressed as  $\gamma_h\simeq 2(L_{j52}/R_{L7}^2)^{1/6}(z_h/R_\star)^{2/9}$, and it is seen that for values of the jet power  inferred from observations, $L_{j52}\simlt1$, the motion of the  head is  sub-to-mildly relativistic inside the stellar envelope.  For the above choice of parameters the head becomes relativistic at $z_h>z_t\simeq 0.014R_\star$. The jet Lorentz factor at the location of the head  is given by $\gamma_j\simeq40(L_{j52}/R_{L7}^2)^{1/14}(z_h/R_\star)^{3/7}$ at $z_h<z_t$, and $\gamma_j\simeq62(L_{j52}/R_{L7}^2)^{1/6}(z_h/R_\star)^{5/9}$ at $z_h>z_t$.  From Eq. (\ref{z_c}) we also have 
\begin{equation}
z_c/R_\star\simeq70\left(\frac{L_{j52}}{R_{L7}^2}\right)^{-3}.
\end{equation}
Thus, the jet will be collimated all the way to the edge of the stellar envelope provided $L_{j52}/R_{L7}^2<4$.   In fact, jets of sufficiently low power may remain confined  well above the edge of the stellar envelope by the surrounding matter that breaks out of the star with the jet.   This raises the question of whether the deconfinement of the jet at breakout is sudden enough to lead to re-acceleration to $\Gamma_j/\theta_j \gg 1$, as proposed recently (Tchekhovskoy et al. 2010; Komissarov et al. 2010).  
The dissipation of the magnetic field in the confinement region may persist after breakout, until the confinement relaxes to the point that current-driven instabilities no longer have time to operate.  From the above results we obtain $\lambda_{CD}/z_h\sim0.55 \beta_A^{-1}(L_{j52}/R_{L7}^2)^{1/3}(z_h/R_\star)^{1/9}$ in the transrelativistic regime ($\gamma_h>1$).  Thus, under the assumption that the cocoon is isobaric it seems that the instability is marginal near and beyond the edge of the star, and it is unclear whether it will indeed become destructive there.  On the other hand, a non-uniform pressure distribution in the cocoon would lead to additional focusing of the jet above the collimation zone and, hence,  a shorter growth length of the instability.  
The hot, low-sigma outflow produced by the destruction of the magnetic field well inside the star and, conceivably, also after breakout will eventually reach the photosphere, and may be the source of the photospheric emission observed during the prompt phase. 

Blazar jets propagating in a medium of constant density, $n_0=\rho_0/m_p$ (measured in cgs units), will be collimated at distances $z>z_c$, with 
\begin{equation}
z_c\simeq 10 \left(\frac{L_{j45}}{n_0 R_{L14}^2}\right)^{3/5}\quad {\rm pc},
\end{equation}
where we adopt  $R_L=10^{14}R_{L14}$ cm.   For $n_0\simeq1$ cm$^{-3}$, $z_c$ is  much larger than the scale over which the magnetic field dissipates, as inferred from observations.  The cocoons observed on tens of kpc scales form, most likely, in the pure hydrodynamic regime (BNPS11).  
However,  the average density in the vicinity of the broad line region may be  larger by several orders of magnitudes than the average ISM density, so that collimation by the cocoon may occur on much smaller scales, shortly after the activation of the central engine.   We note that the Lorentz factor of the confined jet at $z>z_c$ is $\gamma_j\simeq67 (L_{j45}/R_{L14}^2n_0)^{3/10}(z_h/z_c)^{2/9}$.  The latter exceeds the characteristic values inferred from observations, $\gamma_j\sim 10 - 50$, implying that conversion of magnetic energy to kinetic energy can occur by collimation alone, even on sub-parsec scales if the ambient density in the vicinity of the central engine is large enough, viz., $n_0>10^3$.

\section{Conclusion}

We have constructed an analytic model for the propagation of a magnetically dominated jet in an external medium having an arbitrary density profile, and examined the conditions under which the jet can be collimated by the cocoon surrounding it.  Our model assumes that in the region where the jet is well-confined by the cocoon, its structure can be described by the equilibrium solution derived in Lyubarsky (2009) in the equilibrium regime.  This solution is employed as a closure condition for the jet-cocoon system of equations.  We analyzed  two different evolutionary tracks:

The first one, outlined in \S \ref{sec:mag-cocoon}, assumes that some fraction of the magnetic field that enters the jet's head through the reverse shock is advected into the cocoon and remains ordered.  The inner jet is focused by magnetic pinching to a level at which the jet Lorentz factor can be matched to the head through a weak reverse shock.   Under these assumptions a self-consistent solution of the jet-cocoon equations can be obtained.  This solution may correspond to the nose cone revealed in two-dimensional numerical simulations of moderately magnetized jets (Komissarov 1999). However, this scenario requires that the system maintain a high degree of axisymmetry over times longer than the expansion time of the head, which is questionable.  We argued, at the end of \S \ref{sec:mag-cocoon},  that for a dense enough medium such a configuration should be unstable, and should lead to a rapid dissipation of the magnetic field in the inner jet and the surrounding cocoon. 

In the second one, which we find more likely, the magnetic flux in the confined jet dissipates via  current-driven instabilities before it reaches the head, and the jet undergoes a transition from high-to-low sigma above the collimation zone.   A strong reverse shock forms in the low-sigma section of the jet, allowing matching of the jet and head Lorentz factors, as in the pure hydrodynamic case.  Since the side flow that feeds the cocoon is weakly magnetized, magnetic pinching is unimportant. Instead, the jet is confined by the pressure of the gas contained in the inner cocoon.    Collimation commences at a radius at which  the
transverse ram pressure of the unconfined jet roughly equals the cocoon's pressure, followed by spatial oscillations of the confined jet that decay to the equilibrium state by dissipative processes.    The collimation proceeds smoothly if the transverse expansion of the unconfined jet is sub-magnetosonic, or through formation of a (weak)  superfast tangential shock if the transverse expansion is super-magnetosonic.  At the onset of collimation the jet may still be highly magnetized ($\sigma \gg 1$), unlike the pure hydrodynamic case considered in BNPS11.  This leads to scalings different than those derived by BNPS11.  For example, for transrelativistic propagation in an ambient density profile $\rho_a\propto z^{-\alpha}$, we find that the head Lorentz factor evolves according to $\gamma_h\propto z_h^{(3\alpha-2)/18}$  in the high-sigma case, versus  $\gamma_h\propto z_h^{(\alpha-2)/10}$ in the pure hydrodynamic case. 

As long as the jet is well-confined by the cocoon, the growth time of the current-driven instabilities is shorter than the expansion time of the head.  As a consequence, complete destruction of the magnetic field is expected in the confinement region below the head.  The hot, low-sigma matter thereby produced may be a source of high-energy radiation when approaching the photosphere.  In the collapsar scenario for long GRBs, we find that the jet will remain well-confined throughout its propagation in the envelope of the progenitor star, and perhaps even well above the envelope.  For a reasonable density profile, the criterion for the growth of the instability is found to be marginal near the edge of the envelope, so further analysis is required to quantify the likelihood that the instability will become disruptive.  If it does, then magnetic field dissipation may persist for times longer than the duration of the breakout phase.   At any rate, the hot, low-sigma matter produced inside and, conceivably, above the stellar envelope will eventually reach the photosphere and radiate. The photospheric emission observed in the prompt phase of many bursts may be a signature of this mechanism.  Since magnetic field dissipation commences well inside the envelope, at large optical depths, there is sufficient time to generate the radiation entropy required to explain the sub-MeV peaks; typically, a Thomson depth of $\tau>10^3$ is required for complete thermalization (Levinson 2012). %{\bf [Did you estimate how far out the dissipation has to occur to give the observed entropy?  I assume it's not enough if the dissipation occurs at too small z...]}  
The overall shape of the spectrum emitted from the photosphere would depend on the dissipation profile below the photosphere (Levinson 2012; Beloborodov 2012), which in turn depends on the density profile of the progenitor star and other details. 
Further dissipation of the bulk energy of the weakly magnetized fluid near the photosphere may occur via formation of internal or collimation shocks.   

AL acknowledges support from an ISF grant for the Israeli center for high energy astrophysics, and thanks the Fellows of JILA for their hospitality during a sabbatical visit.  MCB acknowledges support from NSF grant AST-0907872 and NASA Astrophysics Theory grant NNX09AG02G.

%%%%%%

\appendix
\section{\label{sec:appA}Oblique MHD shocks in super-fast flows}
The energy momentum tensor of the upstream flow, as measured in the shock frame, is expressed as
\begin{equation}
T_1^{\mu\nu}=(w_1+b_1^2)u_1^\mu u_1^\nu+(p_1+b_1^2/2)g^{\mu\nu}-b_1^\mu b_1^\nu
\end{equation}
where $b_\mu$ is defined below equation (\ref{Tmunu_j}).   We consider a planar shock and choose our coordinate system such that the velocity of the unshocked flow is given by ${\bf \beta}=(\beta_x,0,\beta_z)$, and the shock normal by ${\bf n}=(n_x,0,n_z)$.   For simplicity we assume that magnetic field of the unshocked flow just upstream of the shock is purely toroidal. Then $b_1^\mu=(0,0,b_1,0)$, where $\sqrt{4\pi}b_1=B_{1\phi}/\gamma_1$ is the proper magnetic field.   In terms of the angle $\psi$ between the jet velocity and the shock normal ($\cos\psi={\bf n}\cdot \hat{\bf \beta}$), the jump conditions are written as 

\begin{eqnarray}
\rho_1\gamma_1\beta_1\cos\psi_1=\rho_2\gamma_2\beta_2\cos\psi_2,\label{app-eq-cont}\\
(h_1+\sigma_1)\rho_1\gamma_1^2\beta_1\cos\psi_1=(h_2+\sigma_2)\rho_2\gamma_2^2\beta_2\cos\psi_2\label{app-eq-enrg}\\
(h_1+\sigma_1)\rho_1\gamma_1^2\beta^2_1\cos^2\psi_1+p_1+\rho_1\sigma_1/2=(h_2+\sigma_2)\rho_2\gamma_2^2\beta^2_2\cos^2\psi_2+p_2+\rho_2\sigma_2/2\label{app-eq-momen}\\
\beta_1\sin\psi_1=\beta_2\sin\psi_2\\
\sigma_1/\rho_1=\sigma_2/\rho_2.
\end{eqnarray}
Here subscript $2$ refers to shocked fluid quantities, $\sigma=b^2/\rho$ is the magnetization, and $h=w/\rho$ is the enthalpy per baryon.  Note that $\sqrt{\sigma/h}$ is the Alfven 4-velocity.  Equations (\ref{app-eq-cont}) and (\ref{app-eq-enrg}) can be combined to yield
\begin{equation}
(h_1+\sigma_1)\gamma_1=(h_2+\sigma_2)\gamma_2.\label{enthalpy-cons}
\end{equation}
After some algebraic manipulations, the jump conditions can be reduced to a cubic equation for the variable $x=u_{2\perp}^2=\gamma_2^2\beta_2^2\cos^2\psi_2$  (see also Lyutikov 2004, and Komissarov and Lyutikov  2011, for a similar derivation):
%For a hot upstream flow, $h_1=1+4p_1$, we obtain
\begin{equation}
a_3x^3+a_2x^2+a_1x+a_0=0,\label{cubic}
\end{equation}
with
\begin{eqnarray}
a_3=16[c_1^2-(1-\beta_1^2\sin^2\psi_1)],\\
a_2=16c^2-8c_1c_2\beta_1\cos\psi_1-8(1-\beta_1^2\sin^2\psi_1),\\
a_1=c_2^2\beta_1^2\cos^2\psi_1-8c_1c_2\beta_1\cos\psi_1-(1-\beta_1^2\sin^2\psi_1),\\
a_0=c_2^2\beta_1^2\cos^2\psi_1,
\end{eqnarray}
where  $c_1=\beta_1\cos\psi_1[1+(p_1/(h_1+\sigma_1)+c_2/2)/(u_1^2\cos^2\psi_1)]$
%$c_1=\beta_1\cos\psi_1[1+(1+c_2)/(4u_1^2\cos^2\psi_1)]$ 
and $c_2=\sigma_1/(\sigma_1+h_1)$.  Solutions of (\ref{cubic}) obtained numerically for  $h_1=4p_1/\rho_1$, $\cos\psi_1=0.1$, $\sigma_1/h_1=10$, and different values of the fast magnetosonic Mach number, $M_{f\perp}=u_{1\perp}/u_f$,  $u_f=\sqrt{(h_1+3\sigma_1)/2h_1}$  \footnote{In general, the fast magnetosonic speed $c_f$ satisfies $c_f^2=(\hat{\gamma}p+b^2)/(\rho h+b^2)$, where $\hat{\gamma}$ is the adiabatic index. In the limit $h=4p/\rho$, $\hat{\gamma}=4/3$, it reduces to 
$c_f^2=(h/3+\sigma)/(h+\sigma)$, from which we obtain $u_f^2=c_f^2/(1-c_f^2)=(h+3\sigma)/2h$.} ,  are exhibited in fig \ref{fig:shock}.   As expected, shock solutions exist only for $M_{f\perp}>1$. 
In the limit $\gamma_1>>1$, $M_{f\perp}^2>>1$ Eq. (\ref{cubic}) can be solved analytically to yield,
\begin{eqnarray}
u_{2\perp}^2=\frac{8\sigma^2+10h_1 \sigma+h_1^2}{16h_1(\sigma+h_1)}+\frac{[64\sigma^2(h_1+\sigma)^2+20\sigma(h_1+\sigma)+h_1^4]^{1/2}}{16h_1(\sigma+h_1)},\label{app:sol-jump}
\end{eqnarray}
which in the special case $\cos\psi_1=1$, $h_1=1$ reduces to that obtained by Kennel and Coroniti (1984), and in the limit $\sigma_1=0$, $h_1=1$ reduces to $u_{2\perp}=1/\sqrt{8}$.  Note that $\gamma_2^2=(1+u_{2\perp}^2)/(1-\beta_1^2\sin^2\psi_1)$.  As pointed out by Komissarov (2012), the proper condition for the approximation (\ref{app:sol-jump}) is $M_{f\perp}>>1$, not just $u_{1\perp}>>1$.  In the limit $\sigma_1>>h_1$ the solution (\ref{app:sol-jump}) simplifies to
\begin{eqnarray}
u_{2\perp}^2=(\sigma_1/h_1),\label{sh1}\\
\sin\psi_2=\left(\frac{h_1+\sigma_1}{\sigma_1+h_1\sin^2\psi_1}\right)^{1/2}\sin\psi_1,\\
\frac{\rho_2}{\rho_1}=\frac{\gamma_1\beta_1}{\sqrt{\sigma_1/h_1}}\cos\psi_1,\\
p_{tot}=p_2+b_2^2/2=\frac{1}{4}\left(\sqrt{\frac{h_1+\sigma_1}{\sigma_1}}+1\right)\rho_1h_1\gamma_1^2\beta_1^2\cos\psi_1^2,\label{sh4}
\end{eqnarray}
in agreement with the high Mach number limit of figure \ref{fig:shock}.  In fact, figure \ref{fig:shock} indicates that the latter solution is a good approximation already at modest Mach numbers, $M_{f\perp}=$ a few.  The deflection angle of streamlines across the shock, $\delta=\psi_2-\psi_1$,  is readily obtained from the above:
\begin{equation}
\sin\delta=\left(\frac{\sqrt{1+h_1/\sigma_1}-1}{2\sqrt{\sigma_1/h_1+\sin^2\psi_1}}\right)\sin2\psi_1\simeq \frac{1}{4}(h_1/\sigma_1)^{3/2}\sin2\psi_1.
\end{equation}
The shock compression ration is $r=\beta_1/\beta_2=\beta_1\sqrt{1+h_1/\sigma_1}$, and it is seen that the in the limit $\sigma_1/h_1>>1$ the shock is always weak.  We emphasize that  Eqs.( {\ref{sh1})-(\ref{sh4}) hold not only in the case of hot upstream flow but for any $h_1$.

\begin{figure}[ht]
\centering
\includegraphics[width=12cm]{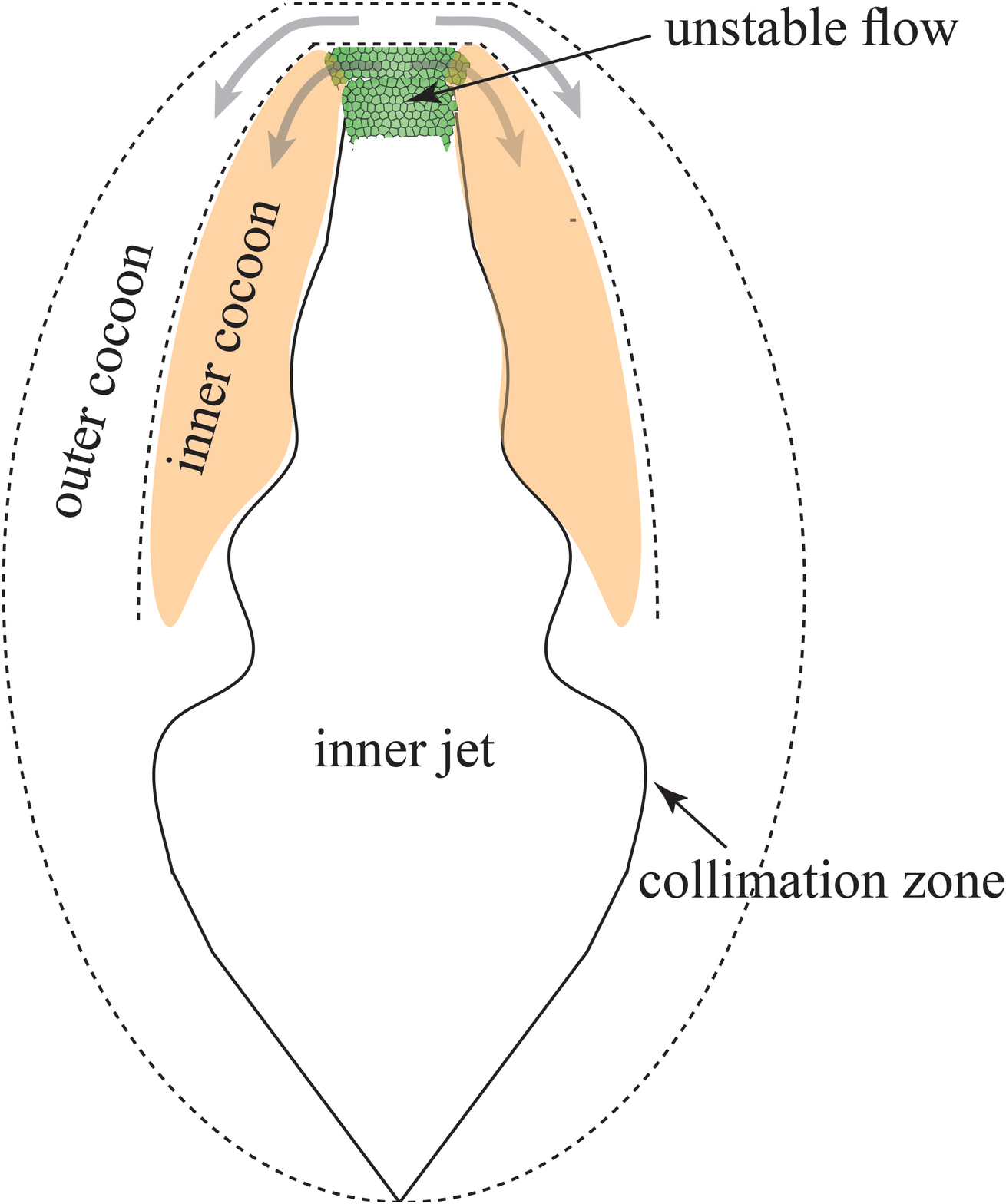}
\caption{\label{fig:schem} Schematic illustration of the jet-cocoon system.}
 \end{figure}

\begin{figure}[ht]
\centering
\includegraphics[width=12cm]{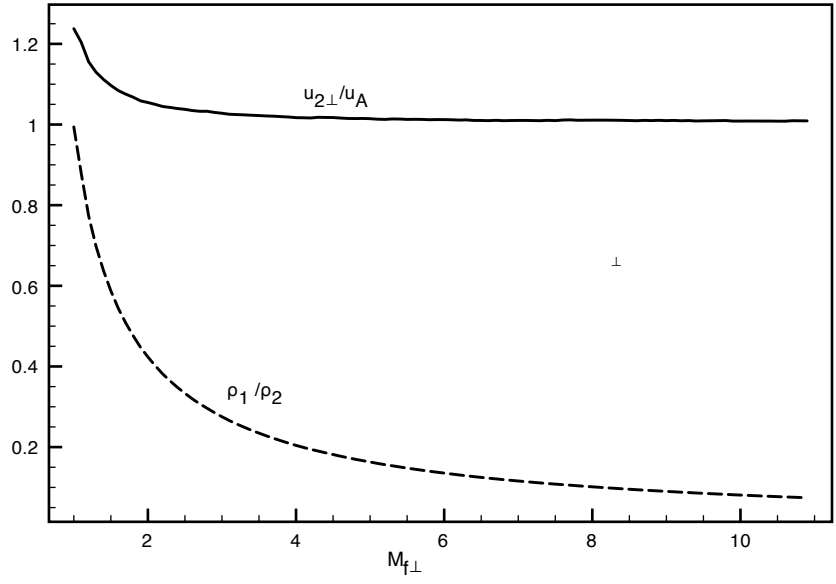}
\caption{\label{fig:shock} Normal component of the downstream 4-velocity normalized to $u_A=\sqrt{\sigma_1/h_1}$ (solid line), and proper density ratio $\rho_1/\rho_2$ (dashed line), versus fast magnetosonic Mach number $M_{f\perp}$, for a hot upstream flow ($h_1=4p_1$) with $\sigma_1/h_1=10$ and incidence angle (measured with respect to the shock normal) $\cos\psi_1=0.1$.}
 \end{figure}

\end{document}